\documentclass[longauth]{aa}

\usepackage{amsmath}
\usepackage{amssymb}

\newcommand{\ra}[4]{${#1}^{\rm h}{#2}^{\rm m}{#3}\fs{#4}$}
\newcommand{\dec}[4]{${#1}\degr{#2}\arcmin{#3}\farcs{#4}$}

\begin{document}

\title{Two new free-floating or wide-orbit planets from microlensing}

\author{
Przemek Mr\'oz\thanks{Corresponding author: pmroz@astrouw.edu.pl}\inst{\ref{warsaw}} \and 
Andrzej~Udalski\inst{\ref{warsaw}} \and 
David~P.~Bennett\inst{\ref{M3},\ref{M13}} \and
Yoon-Hyun Ryu\inst{\ref{kasi}} \and 
Takahiro Sumi\inst{\ref{M1}} \and
Yossi Shvartzvald\inst{\ref{ipac}}, and \and \\
Jan~Skowron\inst{\ref{warsaw}} 
\and Rados\l{}aw~Poleski\inst{\ref{warsaw},\ref{ohio}} \and 
Pawe\l{}~Pietrukowicz\inst{\ref{warsaw}} \and 
Szymon~Koz\l{}owski\inst{\ref{warsaw}} \and 
Micha\l{}~K.~Szyma\'nski\inst{\ref{warsaw}} \and 
\L{}ukasz~Wyrzykowski\inst{\ref{warsaw}} \and 
Igor~Soszy\'nski\inst{\ref{warsaw}} \and 
Krzysztof~Ulaczyk\inst{\ref{warsaw},\ref{warwick}} \and 
Krzysztof~Rybicki\inst{\ref{warsaw}} \and 
Patryk~Iwanek\inst{\ref{warsaw}} (The~OGLE~Collaboration) \and \\
Michael~D.~Albrow\inst{\ref{canterbury}} \and
Sun-Ju~Chung\inst{\ref{kasi},\ref{daejon}} \and
Andrew~Gould\inst{\ref{kasi},\ref{maxplanck},\ref{ohio}} \and
Cheongho~Han\inst{\ref{chungbuk}} \and
Kyu-Ha~Hwang\inst{\ref{kasi}} \and
Youn~Kil~Jung\inst{\ref{kasi}} \and
In-Gu~Shin\inst{\ref{harvard}} \and
Jennifer~C.~Yee\inst{\ref{harvard}} \and
Weicheng~Zang\inst{\ref{tsinghua}} \and
Sang-Mok~Cha\inst{\ref{kasi},\ref{yongin}} \and
Dong-Jin~Kim\inst{\ref{kasi}} \and
Hyoun-Woo~Kim\inst{\ref{kasi}} \and
Seung-Lee~Kim\inst{\ref{kasi},\ref{daejon}} \and
Chung-Uk~Lee\inst{\ref{kasi},\ref{daejon}} \and
Dong-Joo~Lee\inst{\ref{kasi}} \and
Yongseok~Lee\inst{\ref{kasi},\ref{yongin}} \and
Byeong-Gon~Park\inst{\ref{kasi},\ref{daejon}} \and
Richard~W.~Pogge\inst{\ref{ohio}} 
(The KMTNet Collaboration) \and \\
Fumio~Abe\inst{\ref{M2}} \and
Richard~Barry\inst{\ref{M3}} \and
Aparna~Bhattacharya\inst{\ref{M3},\ref{M13}} \and
Ian~A.~Bond\inst{\ref{M6}} \and
Martin~Donachie\inst{\ref{M4}} \and
Akihiko~Fukui\inst{\ref{M5},\ref{M16}} \and
Yuki~Hirao\inst{\ref{M13},\ref{M1}} \and
Yoshitaka~Itow\inst{\ref{M2}} \and
Kohei~Kawasaki\inst{\ref{M1}} \and
Iona~Kondo\inst{\ref{M1}} \and
Naoki~Koshimoto\inst{\ref{M14},\ref{M15}} \and
Man~Cheung~Alex~Li\inst{\ref{M4}} \and
Yutaka~Matsubara\inst{\ref{M2}} \and
Yasushi~Muraki\inst{\ref{M2}} \and
Shota~Miyazaki\inst{\ref{M1}} \and
Masayuki~Nagakane\inst{\ref{M1}} \and
Cl\'ement~Ranc\inst{\ref{M3}} \and
Nicholas~J.~Rattenbury\inst{\ref{M4}} \and
Haruno~Suematsu\inst{\ref{M1}} \and
Denis~J.~Sullivan\inst{\ref{M9}} \and
Daisuke~Suzuki\inst{\ref{M12}} \and
Paul~J.~Tristram\inst{\ref{M10}} \and
Atsunori~Yonehara\inst{\ref{M11}} 
(The MOA Collaboration) \and \\
Dan~Maoz\inst{\ref{wise}} \and
Shai~Kaspi\inst{\ref{wise}} \and
Matan~Friedmann\inst{\ref{wise}}
(The Wise Group)
}


\institute{
Warsaw University Observatory, Al. Ujazdowskie 4, 
00-478 Warszawa, Poland\label{warsaw} \and 
Code 667, NASA Goddard Space Flight Center, Greenbelt, MD 20771, USA\label{M3} \and
Department of Astronomy, University of Maryland, College Park, MD 20742, USA\label{M13} \and
Korea Astronomy and Space Science Institute, Daejon
34055, Republic of Korea\label{kasi} \and 
Department of Earth and Space Science, Graduate School of Science, Osaka University, Toyonaka, Osaka 560-0043, Japan\label{M1} \and
IPAC, Mail Code 100-22, Caltech, 1200 E. California Blvd.,
Pasadena, CA 91125, USA\label{ipac} \and 
Department of Astronomy, Ohio State University, 140 W.
18th Ave., Columbus, OH 43210, USA\label{ohio} \and 
Department of Physics, University of Warwick, Coventry
CV4 7AL, UK\label{warwick} \and
University of Canterbury, Department of Physics and
Astronomy, Private Bag 4800, Christchurch 8020, New Zealand\label{canterbury} \and 
Korea University of Science and Technology, Daejeon 34113,
Republic of Korea\label{daejon} \and 
Max-Planck-Institute for Astronomy, K\"{o}nigstuhl 17,
69117 Heidelberg, Germany\label{maxplanck} \and 
Department of Physics, Chungbuk National University,
Cheongju 28644, Republic of Korea\label{chungbuk} \and 
Harvard-Smithsonian Center for Astrophysics, 60 Garden
St., Cambridge, MA 02138, USA\label{harvard} \and 
Physics Department and Tsinghua Centre for
Astrophysics, Tsinghua University, Beijing 100084, China\label{tsinghua} \and 
School of Space Research, Kyung Hee University,
Yongin, Kyeonggi 17104, Republic of Korea\label{yongin} \and
Institute for Space-Earth Environmental Research, Nagoya University, Nagoya 464-8601, Japan\label{M2} \and
Institute of Natural and Mathematical Sciences, Massey University, Auckland 0745, New Zealand\label{M6} \and
Department of Physics, University of Auckland, Private Bag 92019, Auckland, New Zealand\label{M4} \and
Department of Earth and Planetary Science, Graduate School of Science, The University of Tokyo, 7-3-1 Hongo, Bunkyo-ku, Tokyo 113-0033, Japan\label{M5} \and
Instituto de Astrof\'isica de Canarias, V\'ia L\'actea s/n, E-38205 La Laguna, Tenerife, Spain\label{M16} \and
Department of Astronomy, Graduate School of Science, The University of Tokyo, 7-3-1 Hongo, Bunkyo-ku, Tokyo 113-0033, Japan\label{M14} \and
National Astronomical Observatory of Japan, 2-21-1 Osawa, Mitaka, Tokyo 181-8588, Japan\label{M15} \and
School of Chemical and Physical Sciences, Victoria University, Wellington, New Zealand\label{M9} \and
Institute of Space and Astronautical Science, Japan Aerospace Exploration Agency, 3-1-1 Yoshinodai, Chuo, Sagamihara, Kanagawa, 252-5210, Japan\label{M12} \and
University of Canterbury Mt.\ John Observatory, P.O. Box 56, Lake Tekapo 8770, New Zealand\label{M10} \and
Department of Physics, Faculty of Science, Kyoto Sangyo University, 603-8555 Kyoto, Japan\label{M11} \and
School of Physics and Astronomy and Wise Observatory, 
Tel-Aviv University, Tel-Aviv 6997801, Israel\label{wise}
}

\date{Received 2018 XX XX / Accepted 2018 YY YY}

\titlerunning{Two new free-floating planet candidates from microlensing}
\authorrunning{Mr\'oz et al.}

\abstract{Planet formation theories predict the existence of free-floating planets that have been ejected from their parent systems. Although they emit little or no light, they can be detected during gravitational microlensing events. Microlensing events caused by rogue planets are characterized by very short timescales $t_{\rm E}$ (typically below two days) and small angular Einstein radii $\theta_{\rm E}$ (up to several $\mu$as). Here we present the discovery and characterization of two ultra-short microlensing events identified in data from the Optical Gravitational Lensing Experiment (OGLE) survey, which may have been caused by free-floating or wide-orbit planets. OGLE-2012-BLG-1323 is one of the shortest events discovered thus far ($t_{\rm E}=0.155\pm0.005$\,d, $\theta_{\rm E}=2.37\pm 0.10$\,$\mu$as) and was caused by an Earth-mass object in the Galactic disk or a Neptune-mass planet in the Galactic bulge. OGLE-2017-BLG-0560 ($t_{\rm E}=0.905\pm0.005$\,d, $\theta_{\rm E}=38.7\pm 1.6$\,$\mu$as) was caused by a Jupiter-mass planet in the Galactic disk or a brown dwarf in the bulge. We rule out stellar companions up to a distance of 6.0 and 3.9\,au, respectively. We suggest that the lensing objects, whether located on very wide orbits or free-floating, may originate from the same physical mechanism. Although the sample of ultrashort microlensing events is small, these detections are consistent with low-mass wide-orbit or unbound planets being more common than stars in the Milky Way.}

\keywords{Planets and satellites: detection, Gravitational lensing: micro}

\maketitle

\section{Introduction}

Theories of planet formation predict the existence of free-floating (rogue) planets that are not gravitationally tethered to any host star. These objects could have formed in protoplanetary disks around stars, as ``ordinary'' planets, and could have been ejected as a result of various mechanisms, including planet-planet dynamical interactions (e.g., \citealt{rasio,marzari,marzari2002,chatterjee,scharf2009,veras2009}), ejections from multiple-star systems (e.g., \citealt{kaib2013,sutherland2016}), stellar flybys (e.g., \citealt{malmberg2011,boley2012,moeckel2012}), dynamical interactions in stellar clusters (e.g., \citealt{hurley2002,giersz2009,parker2012,hao2013,liu2013}), or the post-main-sequence evolution of the host star(s) (e.g., \citealt{veras2011, veras2016, kratter2012, voyatzis2013}).

It is believed that low-mass planets are more likely to be scattered to wide orbits or ejected than giant, Jupiter-mass planets. Calculations of \citet{mao2016}, which are based on the core accretion theory of planet formation, predict that most free-floating planets should be of Earth mass. Rogue planets are more likely to form around FGK-type stars, because they are scattered into wide orbits following close encounters with gas giant planets, which are more likely to form around massive stars. The typical total ejected mass is about $5-20\,M_{\oplus}$ and about 10-20\% of all planetary systems should give rise to rogue planets. Similarly, \citet{barclay2017}, using N-body simulations of terrestrial planet formation around solar-type stars, estimated that about 2.5 terrestrial-mass planets are ejected per star in the Galaxy during late-stage planet formation, but these numbers strongly depend on the adopted initial conditions.

Free-floating planetary-mass objects can also be formed by the fragmentation of gas clouds, in a way similar to that in which stars form. Star formation processes are believed to extend down to $1-4\,M_{\rm Jup}$ \citep{boyd2005,whit2006}. This parameter space cannot be probed with the current surveys of young stellar clusters and star-forming regions, which are unable to detect objects less massive than $5-6\,M_{\rm Jup}$ \citep{pena2012,lodieu2013,muzic2015}. Free-floating planetary-mass objects may also form from small molecular cloudlets that have been found in H\,II regions, although it is unclear whether these clouds may contract \citep{gahm2007,grenman2014}.

Gravitational microlensing is the only method that enables us to find Earth-mass free-floating planets. A gravitational microlensing event occurs when a lens (free-floating planet or star) is very closely aligned with a distant source star, with the angular separation smaller than the Einstein radius of the lens $\theta_{\rm E}=5\,\mu\textrm{as}\,\sqrt{M/10\,M_{\oplus}}\,\sqrt{\pi_{\rm rel}/0.1\,\mathrm{mas}}$ (here, $M$ is the lens mass, $\pi_{\rm rel}=\pi_{\rm l}-\pi_{\rm s}$ is the relative lens-source parallax, and $\pi_{\rm l}$ and $\pi_{\rm s}$ are parallaxes to the lens and source, respectively). The gravitational field of the lens can focus light rays of the source, causing a transient brightening of the source to an Earth-based observer. As typical lens-source proper motion in the direction of the Galactic center is $\mu_{\rm rel}=5$\,mas/yr, timescales of microlensing events due to Earth-mass lenses are very short $t_{\rm E} = \theta_{\rm E} / \mu_{\rm rel} \approx $ $10^{-3}\,\mathrm{yr}\approx 0.4\,$d.

Because angular radii of giant source stars in the Galactic bulge $\rho_* = 6\,\mu\mathrm{as}(R/10\,R_{\odot})\,(\pi_{\rm s}/0.125\,\mathrm{mas})$ are comparable to angular Einstein radii of planetary-mass lenses, light curves of giant-source events attributed to free-floating planets should exhibit strong finite source effects (as each point on the source surface is magnified by a different amount). Detection of the finite source effects in the light curve allows us to measure $\theta_{\rm E}$, which can place additional constraints on the mass of the lens. Direct calculations of the lens mass require additional information on $\pi_{\rm rel}$, but the parallax measurement is challenging for such short events (see Introduction in \citealt{Mroz2018}).

Microlensing events on timescales shorter than 2\,d have been traditionally attributed to unbound planets. A statistical analysis of 474 events discovered by the Microlensing Observations in Astrophysics (MOA) group led to the claim of an excess of events on timescales of 1--2\,d (corresponding to Jupiter-mass lenses) and the suggestion that they are caused by a sizable population of Jupiter-mass wide-orbit or free-floating planets \citep{sumi2011}. The analysis of a larger data set collected during the years 2010--2015 of the fourth phase of the Optical Gravitational Lensing Experiment (OGLE-IV) did not confirm these findings \citep{Mroz2017}. \citet{Mroz2017} found a 95\% upper limit on the frequency of Jupiter-mass rogue planets in the Milky Way of 0.25 per star. They detected, however, a few very short events ($t_{\rm E}<0.5$\,d), which could be attributed to Earth- and super-Earth-mass free-floating planets. Their sampling was, however, insufficient to detect finite source effects. See \citet{Mroz2017} and \citet{Mroz2018} for a detailed discussion.

The only known ultrashort microlensing event that exhibited prominent finite source effects, OGLE-2016-BLG-1540, was identified by \citet{Mroz2018} in the OGLE data from the 2016 observing season. This event was likely caused by a Neptune-mass free-floating planet, as inferred from the measurement of the angular Einstein radius. Encouraged by this discovery, we searched for short-timescale microlensing events in the OGLE data from the 2017 season and complemented them with photometric observations from the Korea Microlensing Telescope Network (KMTNet). We also searched for short-duration microlensing events with giant sources in the archival OGLE data collected during the 2010--2015 period.

Here we report the discovery and characterization of two microlensing events, OGLE-2012-BLG-1323 and OGLE-2017-BLG-0560, which can be attributed to free-floating planets. We show that, although the sample of these events is small, these detections are consistent with terrestrial-mass wide-orbit or unbound planets being more common than stars in the Milky Way.

\begin{figure}
\resizebox{\hsize}{!}{\includegraphics{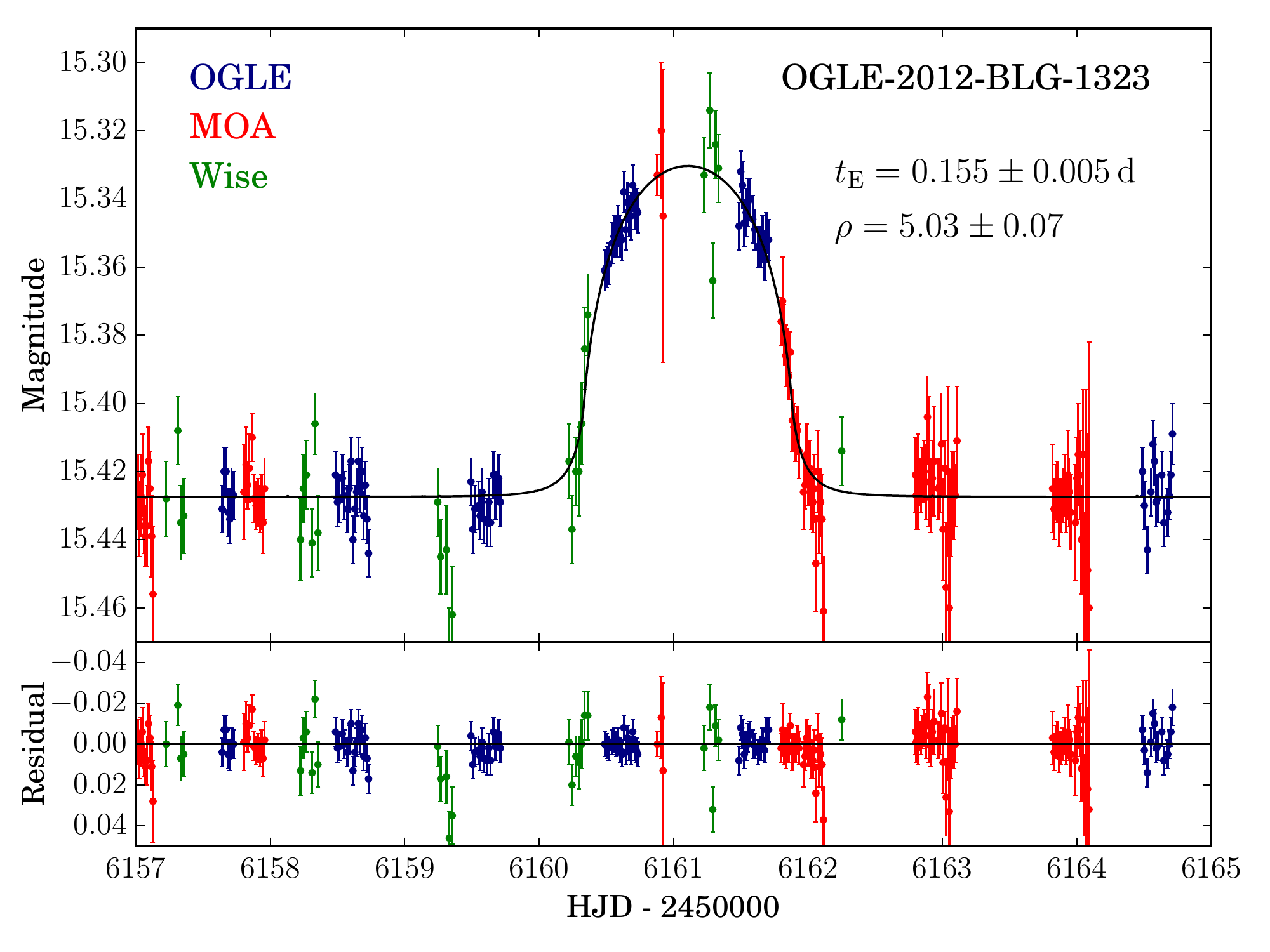}}\\
\resizebox{\hsize}{!}{\includegraphics{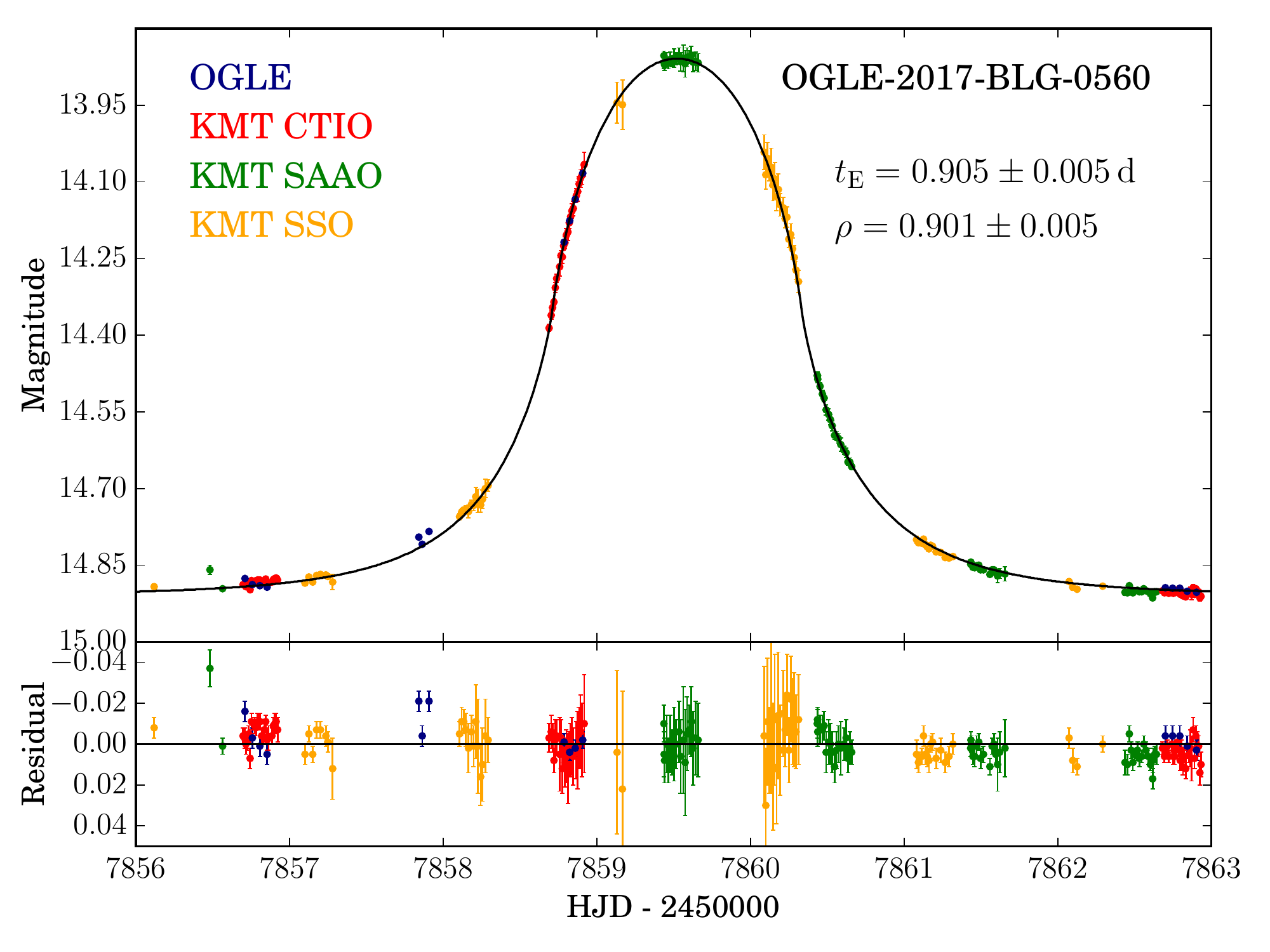}}
\caption{Light curves of two ultrashort microlensing events. Upper panel: OGLE-2012-BLG-1323. Lower panel: OGLE-2017-BLG-0560. Both events show strong finite-source effects, which allows us to measure their angular Einstein radii.}
\label{fig:lc}
\end{figure}

\begin{table*}
\caption{Short-timescale microlensing events exhibiting finite source effects}
\centering
\begin{tabular}{lrrr}
\hline
\hline
Parameter & OGLE-2012-BLG-1323 & OGLE-2017-BLG-0560 & OGLE-2016-BLG-1540\\
\hline
Microlensing model:\\
\hline
$t_0$ (HJD$'$) & $6161.107 \pm 0.008$ & $7859.523 \pm 0.003$ & $7606.726 \pm 0.002$ \\
$t_{\rm E}$ (days) & $0.155 \pm 0.005$ & $0.905 \pm 0.005$ & $0.320 \pm 0.003$ \\
$u_0$  & $0.63 ^{+0.64}_{-0.44}$ & $0.105 ^{+0.031}_{-0.045}$ & $0.53 \pm 0.04$ \\
$\rho$  & $5.03 \pm 0.07$ & $0.901 \pm 0.005$ & $1.65 \pm 0.01$ \\
$I_{\rm s}$ & $15.43 \pm 0.05$ & $14.91 \pm 0.05$ & $14.76 \pm 0.05$\\
$f_{\rm s}$ & $1.00$ (fixed) & $1.00$ (fixed) & $1.00$ (fixed)\\
\hline
Source star:\\
\hline
$I_{\rm S,0}$ & $14.09 \pm 0.06$ & $12.47 \pm 0.05 $ & $13.51 \pm 0.09$\\
$(V-I)_{\rm S,0}$ & $1.73 \pm 0.02$ & $2.31 \pm 0.02$ & $1.67 \pm 0.02$\\
$(V-K)_{\rm S,0}$ & $3.77 \pm 0.03$ & $4.73 \pm 0.06$ & $3.67 \pm 0.03$\\
$T_{\rm eff}$ (K) & $3800 \pm 200$ & $3600 \pm 200$ & $3900 \pm 200$ \\
$\Gamma$ (limb darkening, $I$ band) & 0.40 & 0.41 & 0.36\\
$\Lambda$ (limb darkening, $I$ band) & 0.30 & 0.28 & 0.34\\
$\theta_*$ ($\mu$as) & $11.9 \pm 0.5$ & $34.9 \pm 1.5$ & $15.1 \pm 0.8$\\
\hline 
Physical parameters: \\
\hline
$\theta_{\rm E}$ ($\mu$as) & $2.37 \pm 0.10$ & $38.7 \pm 1.6$ & $9.2 \pm 0.5$\\
$\mu_{\rm rel,geo}$ (mas\,yr$^{-1}$) & $5.6 \pm 0.3$ & $15.6 \pm 0.7$ & $10.5 \pm 0.6$\\
\hline
\end{tabular}
\tablefoot{
HJD$'$=HJD--2450000. $f_{\rm s}=F_{\rm s}/(F_{\rm s}+F_{\rm b})$ is the blending parameter. Parameters for OGLE-2016-BLG-1540 are shown for comparison and are taken from \citet{Mroz2018}.
}
\label{tab:pars}
\end{table*}

\section{Observations}

Microlensing event OGLE-2017-BLG-0560 was announced on 2017 April~16 by the OGLE Early Warning System \citep{udalski2003}. This event was located at equatorial coordinates of R.A. = \ra{17}{51}{51}{33}, Dec. = \dec{-30}{27}{31}{4} (J2000.0) in the field BLG534, which  was observed with a cadence of 60 minutes. The OGLE survey operates from Las Campanas Observatory, Chile, and uses a dedicated 1.3 m Warsaw Telescope, equipped with a mosaic CCD camera with a field of view of 1.4\,deg$^2$ (see \citealt{udalski2015} for details of the survey).

This event was also observed by three identical 1.6 m telescopes from the KMT Network \citep{kim2016}, which are located at the Cerro Tololo Inter-American Observatory (CTIO; Chile), the South African Astronomical Observatory (SAAO; South Africa), and the Siding Spring Observatory (SSO; Australia). The event was located in the two overlapping fields BLG01 and BLG41, each  observed with a cadence of 30 minutes. For the modeling, we used observations collected between March 7 and May 26, 2017.

The second event analyzed in this paper, OGLE-2012-BLG-1323, was also discovered by the OGLE Early Warning System, on 2012 August~21. This event is located at equatorial coordinates of R.A. = \ra{18}{00}{18}{51}, Dec. = \dec{-28}{35}{01}{7} (J2000.0) in the field BLG512, which was monitored with a cadence of 20 minutes. This event was not previously identified as a free-floating planet candidate \citep{Mroz2017}, owing to its extremely low amplitude (below 0.1 mag).

We supplement OGLE observations with the data from the MOA \citep{bond2001} and Wise groups \citep{yossi2016}. MOA observations were collected using the 1.8 m telescope at Mt. John University Observatory in New Zealand \citep{sumi2013}. Wise observations were taken with the 1 m telescope at Wise Observatory in Israel equipped with the LAIWO camera.

All data were taken in the $I$ band except for MOA data; the MOA group uses a custom wide filter, which is effectively the sum of the standard $R$ and $I$ filters. Photometry was extracted using custom implementations of the difference image analysis technique: \citeauthor{wozniak} (\citeyear{wozniak}; OGLE), \citeauthor{albrow2017} (\citeyear{albrow2017}; KMTNet and Wise), and \citeauthor{bond2001} (\citeyear{bond2001}; MOA).

\section{Light curve modeling}
\label{sec:modeling}

Light curves of both events are well described by the extended-source point-lens model (Fig. \ref{fig:lc}), which is defined by four parameters: $t_0$ (time of the closest lens-source approach), $u_0$ (impact parameter in Einstein radius units), $t_{\rm E}$ (event timescale), and $\rho=\theta_*/\theta_{\rm E}$ (normalized radius of the source, i.e., the ratio of the angular radius of the source $\theta_*$ to the angular Einstein radius $\theta_{\rm E}$). Two additional parameters (for each observatory and filter) are needed to describe the source star flux ($F_{\rm s}$) and unmagnified flux of the blend ($F_{\rm b}$). When the blend flux is allowed to vary, the best-fit solutions are characterized by negative blending ($F_{\rm b}<0$). We therefore, following the approach of \citet{Mroz2018}, kept $F_{\rm b}=0$ constant, but we also added in quadrature 0.05 mag to the uncertainty of the source brightness.

The best-fit parameters and their uncertainties are shown in Table \ref{tab:pars}. The uncertainties are estimated using the Markov chain Monte Carlo technique \citep{foreman2013} and represent 68\% confidence intervals of marginalized posterior distributions.

To describe the brightness profile of the source star, we adopted the square-root limb-darkening law, described by two parameters $\Gamma$ and $\Lambda$ (which are filter-dependent; \citealt{yoo2004}). If allowed to vary, $\Gamma$ and $\Lambda$ are strongly correlated. We thus kept limb-darkening coefficients constant, using the limb-darkening models of \citet{claret2011} (see Table \ref{tab:pars} for their numerical values). We used ATLAS models and assumed a solar metallicity, microturbulent velocity of 2\,km/s, and surface gravity of $\log\,g=2.0$ ($\Gamma$ and $\Lambda$ are weakly dependent on $\log\,g$ if $\log\,g\leq 2.0$), as is appropriate for giant sources.

The archival light curve of OGLE-2017-BLG-0560 shows low-amplitude (0.02 mag), semi-regular variability that is typical of OGLE small amplitude red giants \citep{eyer}. The strongest pulsation period in the 2017 data is 18.9\,d. As the effective duration of the event (3 days) is much shorter than the pulsation period, we expected that the inferred model parameters should not be strongly influenced by the variability of the source. Additional modeling, in which we assume that the flux of the source varies sinusoidally with a period of 18.9\,d, results in almost identical microlensing parameters (within the error bars) to those of the model with the constant source. 

We also searched for terrestrial parallax signal \citep{gould2009,yee2009,freeman2015}  in the light curve of OGLE-2012-BLG-1323, but the $\chi^2$ improvement was insignificant ($\Delta\chi^2=1$) and the limits on the microlens parallax were very poor. We did not fit the parallax model to the light curve of OGLE-2017-BLG-0560 because of the low-level variability of the source.

Finally, we also searched for possible binary lens models. Short-duration events may be caused by close binary lenses (when the projected separation $s$, in Einstein radius units, is much smaller than 1), when the source crosses a small triangle-shaped caustic that is far ($\sim 1/s$) from the center of mass. The expected light curves are asymmetric, unless the source is larger than the caustic. In that case the light curve may superficially look like an extended-source point-lens event, except that it has a more extended tail. We found that the best-fitting close binary models are disfavored by $\Delta\chi^2$ of several hundred for OGLE-2012-BLG-1323 and even more for OGLE-2017-BLG-0560. The latter event has a large amplitude ($\approx 1$\,mag), but the peak magnification in close binary models is usually much lower than that, unless the source is small ($\rho < 0.001$) and the light curve is asymmetric. We cannot rule out that the lens is a wide-orbit planet; we discuss these cases in Section \ref{sec:wide}.

\section{Physical parameters}
\label{sec:physical}

\subsection{Source stars}

\begin{figure*}
\centering
\includegraphics[width=8.8cm]{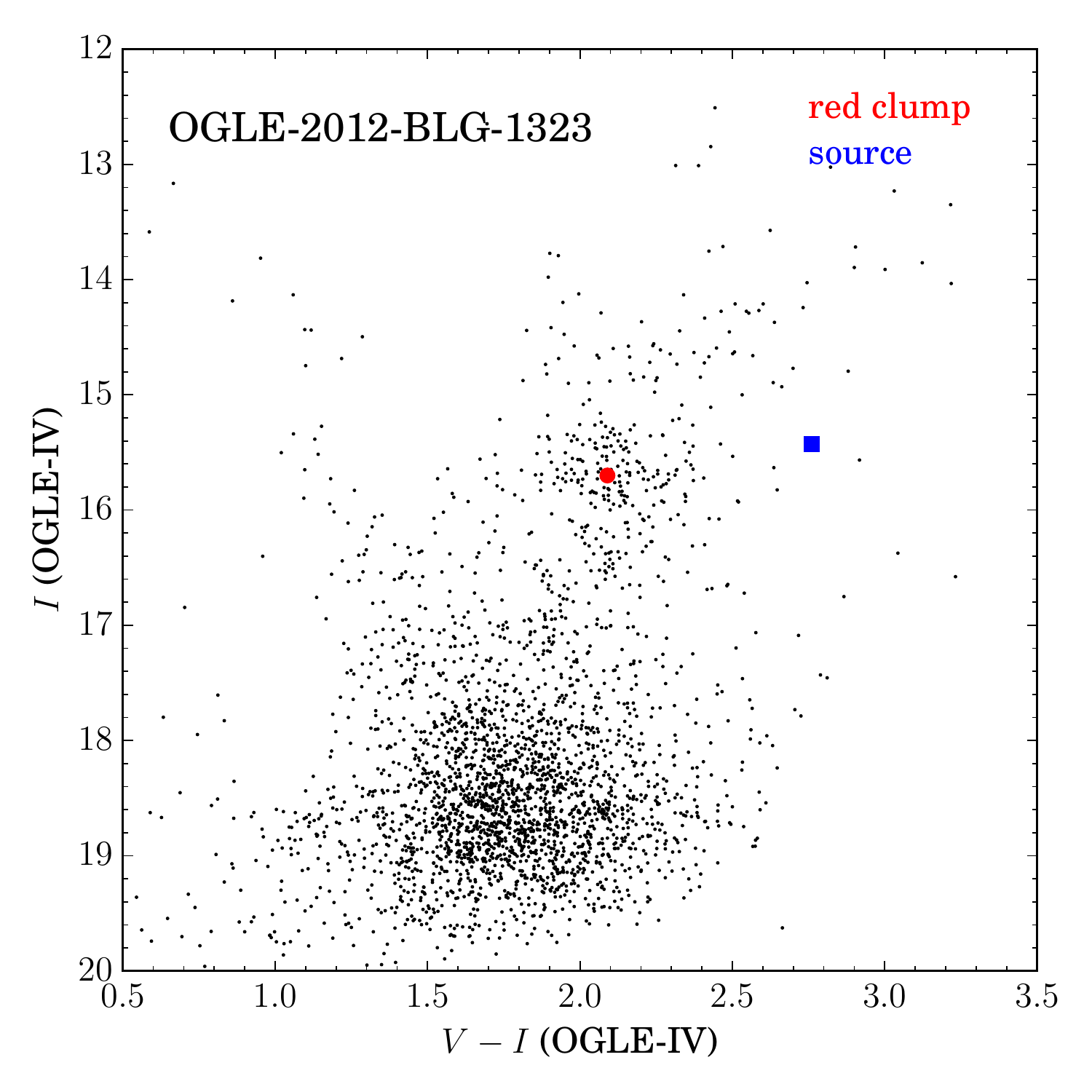}
\includegraphics[width=8.8cm]{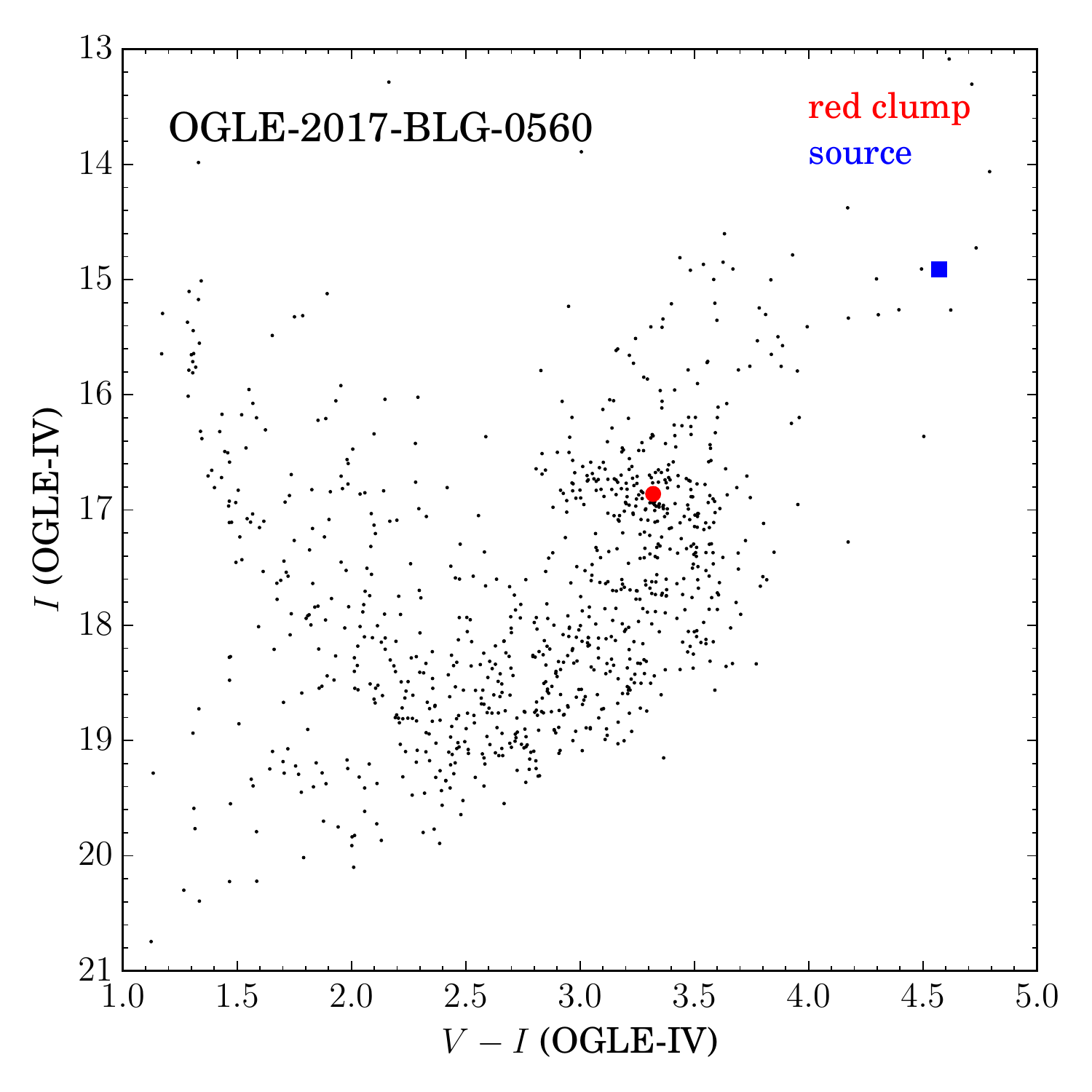}
\caption{OGLE-IV color--magnitude diagrams for stars in $2'\times 2'$ regions around OGLE-2012-BLG-1323 and OGLE-2017-BLG-0560. Sources are marked with blue squares and are likely located in the Galactic bulge. If sources were M dwarfs, they would have absolute $I$-band magnitudes of 9.75 (OGLE-2012-BLG-1323) and 13.90 (OGLE-2017-BLG-0560) \citep{pecaut2013} and would be located at a distance of 140\,pc and 16\,pc, respectively, which contradicts the \textit{Gaia} DR2 parallaxes ($0.15 \pm 0.14$\,mas and $0.23 \pm 0.19$\,mas, respectively).}
\label{fig:cmd}
\end{figure*}

Model parameters can be translated into physical parameters of the lens provided that the angular radius of the source star is known. Here we use a standard technique \citep{yoo2004} of measuring the offset of the source from the centroid of red clump giants in the calibrated color--magnitude diagram in a $2'\times 2'$ region around the event (about $5\,\mathrm{pc}\times5\,\mathrm{pc}$ at the Galactic center distance; Figure~\ref{fig:cmd}). Because we lack color observations collected during the two events, our best estimate for the color of the source is the color of the baseline star. This is further supported by the lack of evidence for blending in the $I$-band light curves and the low probability of  bright unmagnified blends. As the intrinsic color \citep{bensby2011} and dereddened brightness of the red clump \citep{nataf2013} are known toward a given direction, we are able to calculate the dereddened color and brightness of the source. Subsequently, we use color--color \citep{bessell1988} and color--surface brightness (CSB) \citep{kervella2004} relations for giants to measure the angular radius of the source star\footnote{As both sources are very red, it is important to determine how well the empirical CSB relations are calibrated in this range. The relation of \citet{kervella2004} was derived for giants with colors $0.9<(V-K)_0<2.5$, but it agrees well with the earlier relation by \citet{fouque1997}, which is valid in a wider color range. \citet{groene2004} published a CSB relation for M giants ($3.2<(V-K)_0<6.1$), which gives angular radii that are systematically 10\% lower than those based on \citet{kervella2004}: $\theta_*=10.9\pm 0.7$\,$\mu$as (OGLE-2012-BLG-1323) and $\theta_*=29.8\pm 1.9$\,$\mu$as (OGLE-2017-BLG-0560). \citet{adams2018} recently published a new CSB relation for giants ($-0.01<(V-I)_0<1.74$), from which we find $\theta_*=11.5\pm 0.9$\,$\mu$as (OGLE-2012-BLG-1323) and $\theta_*=32.3\pm 2.3$\,$\mu$as (OGLE-2017-BLG-0560), in good agreement with our determination.}. We also use the color--temperature relations of \citet{houdashelt2000,hou2000} and \citet{ramirez2005} to estimate the effective temperature of the source. The angular Einstein radius is $\theta_{\rm E}=\theta_{*}/\rho$ and the relative lens-source proper motion (in the geocentric frame) is $\mu_{\rm rel,geo}=\theta_{\rm E} / t_{\rm E}$. The heliocentric correction ($v_{\oplus,\perp}\pi_{\rm rel}/\mathrm{au}$, where $v_{\oplus,\perp}$ is the Earth's velocity projected on the sky), which should be added vectorially, is of the order of $3 \pi_{\rm rel}\,\mathrm{yr}^{-1}$ and is negligible unless the lens is nearby (closer than 1\,kpc from the Sun). The physical parameters of the source star and lens are given in Table~\ref{tab:pars}.

\begin{figure*}
\centering
\includegraphics[width=8.8cm]{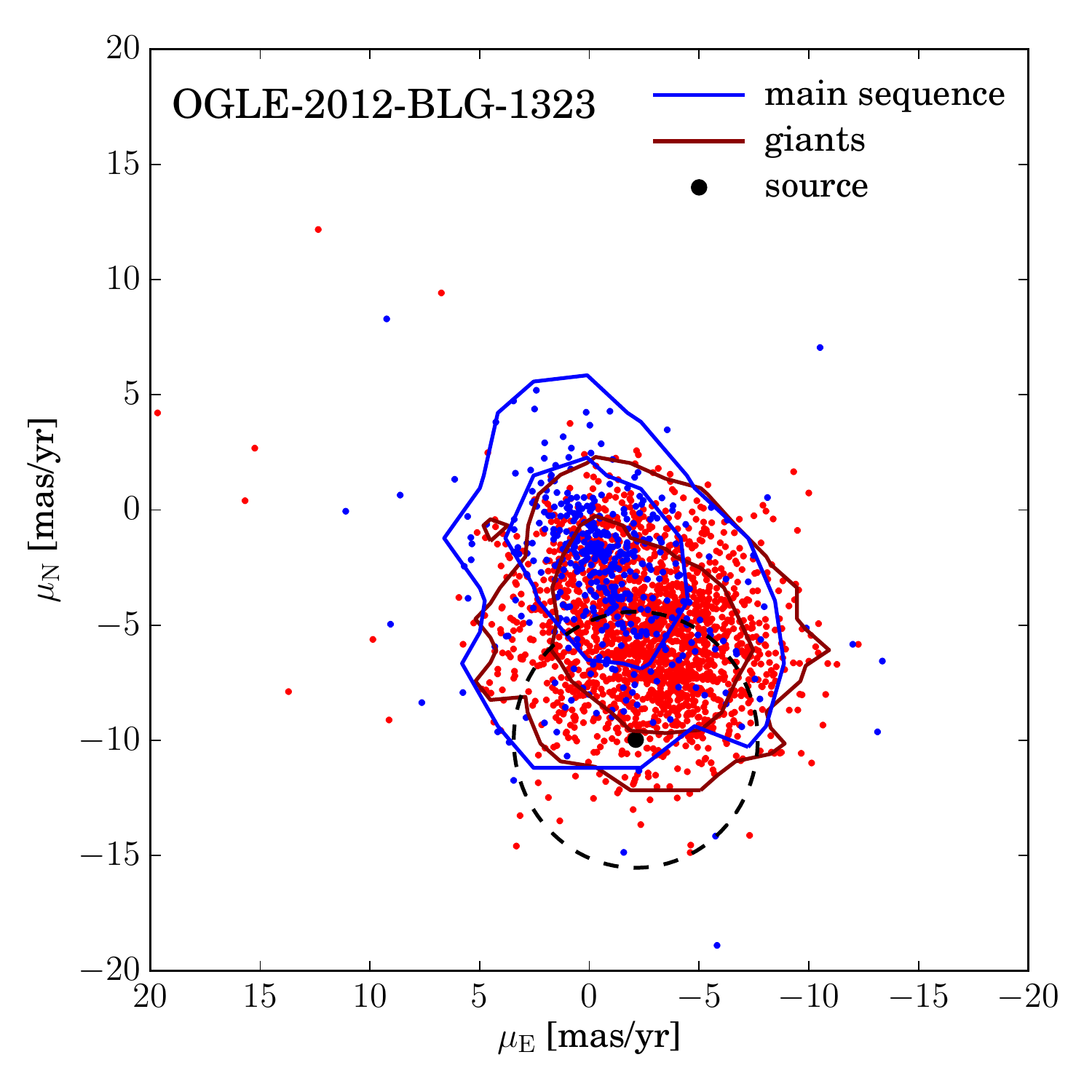}
\includegraphics[width=8.8cm]{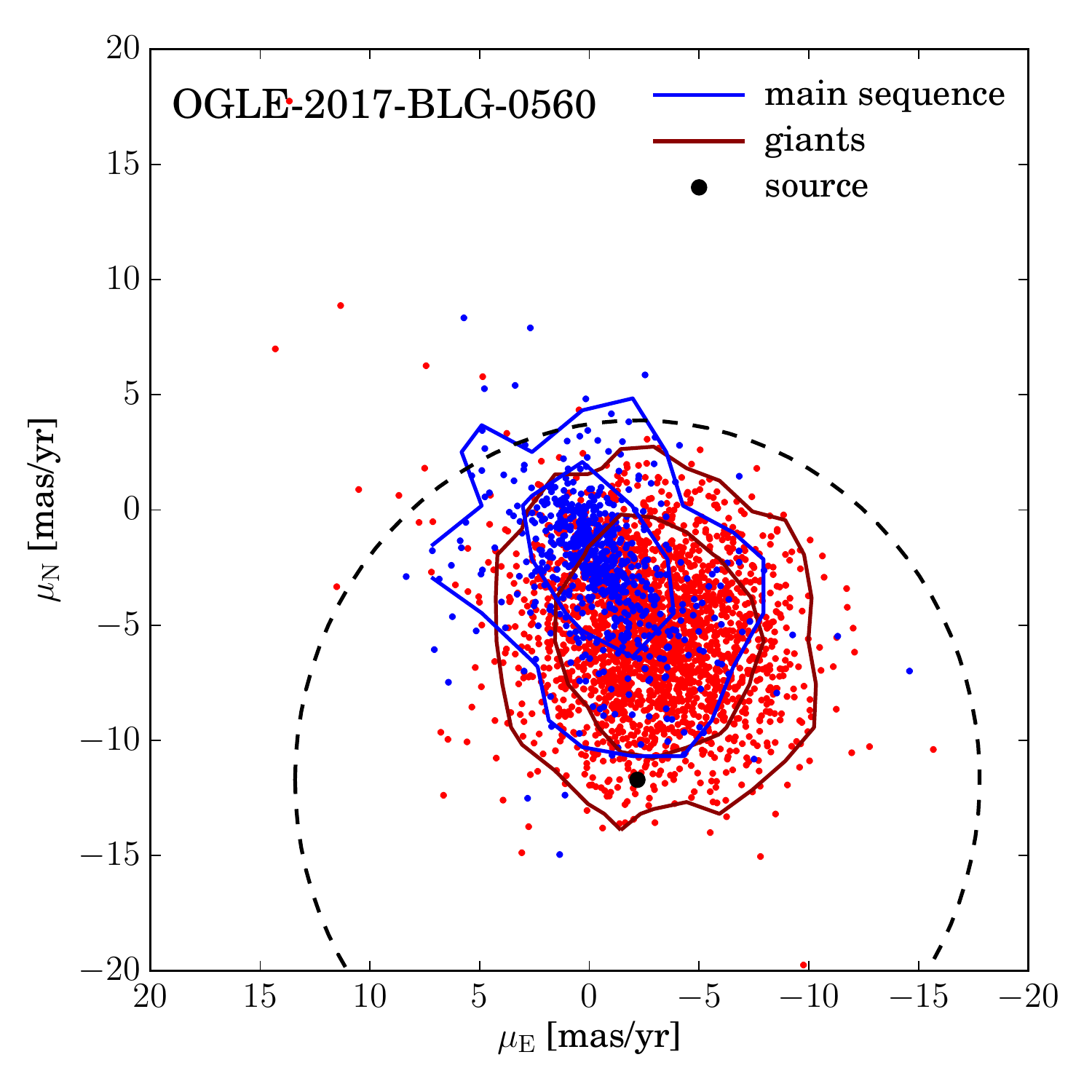}
\caption{\textit{Gaia} DR2 proper motions of stars within $4'$ of OGLE-2012-BLG-1323 (left) and OGLE-2017-BLG-0560 (right). Blue contours correspond to the main-sequence stars (Galactic disk population) and red contours to giants (bulge population). Solid contours enclose 68\% and 95\% of all objects. The source is marked with a black dot. The black dashed circle corresponds to the relative source-lens proper motion of 5.6\,mas/yr (left) and 15.6\,mas/yr (right).}
\label{fig:pm}
\end{figure*}

\subsection{Proper motion of source stars}

Because source stars are bright and lenses contribute little (if any) light, the absolute proper motions of the sources can be found in the second \textit{Gaia} data release (DR2) \citep{gaia2016,gaia2018}. We recall, however, that the \textit{Gaia} performance in the crowded regions of the Galactic center is poor, especially for faint sources. Figure \ref{fig:pm} shows proper motions of stars located within $4'$ of the sources. In both cases source proper motions are consistent with those of Galactic bulge stars (represented by red clump and red giant branch stars), although proper motions measured relative to the mean velocity of the bulge are high. This contributes to the high relative lens-source proper motion of OGLE-2017-BLG-0560 and its very short timescale.

\subsection{Constraints on the host star}
\label{sec:wide}

\begin{figure}
\resizebox{\hsize}{!}{\includegraphics{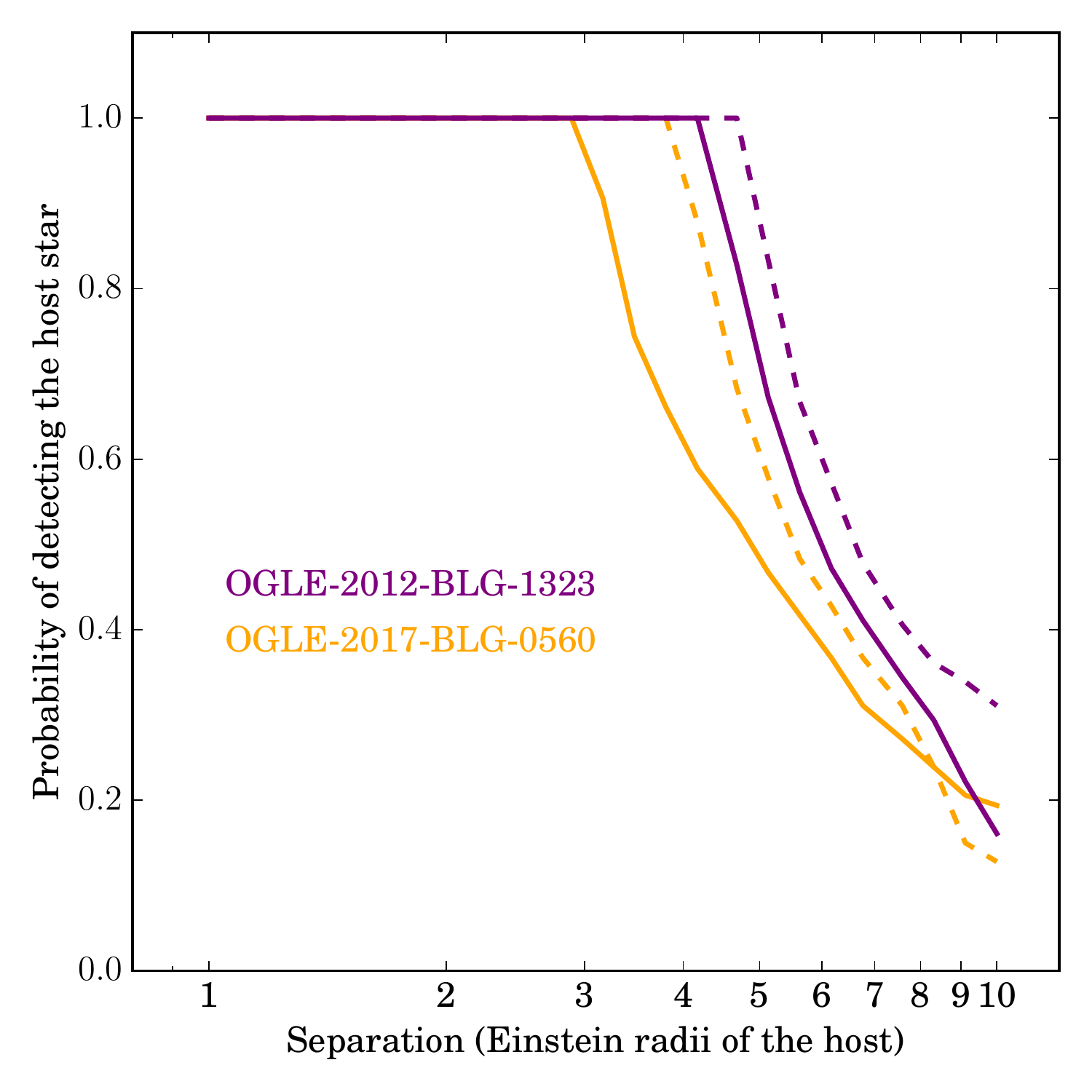}}
\caption{Probability of detecting the putative host star as a function of star-planet separation. Solid curves correspond to the lens located in the Galactic bulge ($\pi_{\rm rel}=0.01$\,mas) and dashed curves to the lens in the Galactic disk ($\pi_{\rm rel}=0.1$\,mas).} 
\label{fig:limits}
\end{figure}

If the trajectory of the source passed near a putative host star, we would detect additional anomalies in the light curves of both events. As we have not found any, we are able to provide only lower limits on the projected star-planet separation, using the method of \citet{Mroz2018}. In short, the description of a binary lens requires three additional parameters: mass ratio $q$, separation $s$ (in Einstein radius units), and angle $\alpha$ between the source trajectory and binary axis. We consider a $0.3\,M_{\odot}$ host located either in the Galactic disk ($\pi_{\rm rel}=0.1$\,mas) or in the bulge ($\pi_{\rm rel}=0.01$\,mas), which corresponds to $\theta_{\rm E,host}=0.49$\,mas or $0.16$\,mas, respectively. Then, for each pair of mass ratio $q=\sqrt{\theta_{\rm E}/\theta_{\rm E,host}}$ and separation $s$, we simulate 180 OGLE light curves (spanning from 2010 March 4 through 2017 October 30) with uniformly distributed $\alpha$, and calculate the fraction of light curves that show signatures of the putative host star (see Fig. \ref{fig:limits}). For OGLE-2012-BLG-1323 we find a 90\% lower limits of 11.8\,au for the disk case (4.9 Einstein radii of the host) and 6.0\,au for the bulge host (4.5 Einstein radii of the host). The formal 90\% limits for OGLE-2017-BLG-0560 are 9.3\,au and 3.9\,au, respectively, but the sensitivity to additional anomalies in the light curve is reduced, owing to low-level variability of the source.

\section{Discussion}
\label{sec:discussion}

The two microlensing events presented in this paper and OGLE-2016-BLG-1540 \citep{Mroz2018} share a number of similarities (Table \ref{tab:pars}). All events occurred on bright giant stars (with estimated angular radii of $9.2-34.9$\,$\mu$as) and their relative lens-source proper motions are high ($5.6-15.6$\,mas/yr). All three events show prominent finite source effects, which led to the measurement of the angular Einstein radius. The fact that all three events occurred on bright, large sources is surprising as less than 3\% of all known events are found with sources brighter than $I=16$\,mag. Moreover, the microlensing event rate $\Gamma$ is proportional to the area of the sky swept by the Einstein ring: $\Gamma\propto \theta_{\rm E}\mu_{\rm rel}$. The high lens-source relative proper motion makes an event more likely to be found, but events with $\mu_{\rm rel}>10\,$mas\,yr$^{-1}$ are very rare \citep{han2017}.

Strong finite source effects make the duration of an event longer, especially if $\rho \gg 1$, which makes giant source events easier to detect. The typical timescale $t_*$ of such an event is comparable to the time needed for the lens to cross the chord of the source:
\begin{equation}
t_* = 2 t_{\rm E} \rho \sqrt{1-\left(\frac{u_0}{\rho}\right)^2}.
\label{eq:teff}
\end{equation}
Monitoring of giant-star microlensing events, as advocated by the Hollywood strategy of \citet{gould1997}, is therefore a promising way of studying free-floating planets.

On the other hand, the peak magnification $A_{\rm peak}$ in the absence of blending declines with the source size \citep{gg1997}:
\begin{equation}
A_{\rm peak} \approx 1+\frac{2}{\rho^2}\,(\mathrm{for}\rho\gg 1).
\end{equation}
The OGLE Early Warning System alerts events that brighten by at least 0.06~mag \citep{udalski2003}, which corresponds to $\rho\lesssim6$, but the search algorithm of \citet{Mroz2017} is sensitive to lower magnifications. Equation (\ref{eq:teff}) also explains why the impact parameter of OGLE-2012-BLG-1323 is poorly measured (cf. Table \ref{tab:pars}): for large sources ($\rho=5$) changing the impact parameter from $u_0=0$ to $u_0=1$ leads to an increase in $t_{\rm E}$ of only 2\%, which is already included in the reported uncertainties.

\begin{table*}
\caption{Posterior distributions for the lens mass calculated from the Bayesian analysis}
\centering
\begin{tabular}{lccccc}
\hline
\hline
Percentile              & 2.3th  &   15.9th  &  50.0th  &  84.1th  &   97.7th\\
\hline

Prior 1: \\

OGLE-2012-BLG-1323 & $1.2 M_{\oplus}$ & $5.0 M_{\oplus}$ & $12 M_{\oplus}$  & $47 M_{\oplus}$   & $480 M_{\oplus}$\\
OGLE-2017-BLG-0560 & $5.7 M_{\rm Jup}$ & $24 M_{\rm Jup}$  & $100 M_{\rm Jup}$ & $0.40 M_{\odot}$ & $0.75 M_{\odot}$\\
OGLE-2016-BLG-1540 & $22 M_{\oplus}$ & $110 M_{\oplus}$ & $1.5 M_{\rm Jup}$ & $47 M_{\rm Jup}$ & $58 M_{\rm Jup}$ \\

\hline

Prior 2: \\
OGLE-2012-BLG-1323 & $3.8 M_{\oplus}$ & $7.8 M_{\oplus}$ & $30 M_{\oplus}$  & $240 M_{\oplus}$  & $990 M_{\oplus}$\\
OGLE-2017-BLG-0560 & $1.1 M_{\rm Jup}$ & $13 M_{\rm Jup}$  & $77 M_{\rm Jup}$  & $0.35 M_{\odot}$ & $0.74 M_{\odot}$\\
OGLE-2016-BLG-1540 & $32 M_{\oplus}$ & $120 M_{\oplus}$ & $2 M_{\rm Jup}$ & $40 M_{\rm Jup}$ & $53 M_{\rm Jup}$ \\

\hline
\end{tabular}
\tablefoot{
We use the Galactic model of \citet{bennett2014} and the planetary mass function $dN/dM\propto M^{-1.8}$ (prior~1) or $dN/dM\propto M^{-1.3}$ (prior~2).
}
\label{tab:bayes}
\end{table*}

The mass of the lens depends on the angular Einstein radius $\theta_{\rm E}$ and the relative parallax $\pi_{\rm rel}$:
\begin{equation}
M = \frac{\theta_{\rm E}^2}{\kappa\pi_{\rm rel}}.
\end{equation}
For both events, the masses cannot be unambiguously determined because the microlens parallaxes cannot be measured. For lenses located in the Galactic disk ($\pi_{\rm rel}\approx0.1$\,mas) the masses are $2.3\,M_{\oplus}$ and $1.9\,M_{\rm Jup}$ for OGLE-2012-BLG-1323 and OGLE-2017-BLG-0560, respectively. If the lenses are located in the Galactic bulge ($\pi_{\rm rel}\approx0.01$\,mas), they have higher masses of about $23\,M_{\oplus}$ or $20\,M_{\rm Jup}$, respectively.

The mass of the lens can be further constrained by employing the Bayesian analysis. If we assume a Galactic model and a mass function of lenses, we can estimate the mass of the lens that most likely reproduces the observations (Einstein radius and proper motion) given the model. However, as we probe extreme events, we must be cautious that the inferred masses will depend on priors that may be derived from data on objects whose properties lie well beyond the range of those being probed. We used the Galactic model of \citet{bennett2014} and two planetary mass functions as the prior. The first mass function ($dN/dM\propto M^{-1.3}$), taken from \citet{sumi2011}, overpredicts the number of short-timescale events \citep{Mroz2017}. The second function ($dN/dM\propto M^{-1.8}$) is steeper, which is consistent with the findings of \citet{clanton2016} and \citet{suzuki2016}, and better describes the event timescale distribution of \citet{Mroz2017}. The posterior distributions favor Galactic bulge lenses, but the allowed range of masses is very broad (see Table~\ref{tab:bayes}). The median masses of the Bayesian results,
$$
M = (0.038,1.5,100)\,M_{\rm Jup},
$$
when combined with the angular Einstein radius values in Table~\ref{tab:pars},
$$
\theta_{\rm E} = (2.37,9.2,38.7)\,\mu{\rm as},
$$
correspond to lens-source relative parallaxes of 
$$
\pi_{\rm rel} = (0.019,0.007, 0.002)\,\rm{mas}
$$
for the three events OGLE-2012-BLG-1323, OGLE-2016-BLG-1540, and OGLE-2017-BLG-0560, respectively.

The ultrashort timescale event OGLE-2012-BLG-1323 ($t_{\rm E}=0.155\pm0.005$\,d) is almost certainly caused by a planetary-mass object (Earth- to super-Earth-mass), while the mass of OGLE-2016-BLG-1540 ($t_{\rm E} = 0.320 \pm 0.003$\,d) is poorly constrained. The rate of events due to brown dwarfs and stars with timescales shorter than 0.32\,d (0.155\,d) is just $10^{-5}$ ($5\times 10^{-7}$) of the total event rate \citep{Mroz2017}.

It is not possible to determine, without further high-resolution follow-up observations, whether  these planets are free-floating or are located at very wide orbits. Owing to their high relative lens-source proper motions, such searches will be possible in the near future with current instruments or next-generation telescopes \citep{gould2016}. As the sources are bright, separations of $\sim 100$\,mas are required to resolve the putative host stars; such separations will be reached in the late 2020s.

Presently, there are no observational constraints on the frequency of bound Earth- and super-Earth-mass wide-orbit planets as their detection is challenging with the current techniques. For example, \citet{poleski2014} found a $4\,M_{\rm Uranus}$ planet at projected separation of 5.3 Einstein radii and \citet{sumi2016} discovered a Neptune analog at projected separation of 2.4 Einstein radii. Planet-formation theories, such as the core accretion model \citep{ida2004}, predict very few low-mass planets at wide orbits because the density of solids and gas in a protoplanetary disk is very low at such large separations. It is believed that Uranus and Neptune formed closer to the Sun, near Jupiter and Saturn, and were subsequently scattered into wide orbits \citep{thommes1999,thommes2002}. Likewise, multiple protoplanets of up to a few Earth masses can be scattered to wide orbits and eventually ejected by growing gas giants (e.g., \citealt{chatterjee,izidoro2015,bromley2016,silsbee2018}). From the point of view of microlensing observations, these objects, whether bound or free-floating, are in practice indistinguishable.

While making statistical inferences out of such a small sample of events is risky, we show that these detections are consistent with low-mass lenses being common in the Milky Way, unless it is just a coincidence that the  events occurred on bright giant stars. According to models presented by \citet{Mroz2017}, about $2.8\times 10^{-3}$ of all events should be caused by Earth- and super-Earth-mass lenses (on timescales $t_{\rm E} < 0.5$\,d) if there were one such object per each star. About 50 events with giant sources brighter than $I=16$ are found in OGLE high-cadence fields annually, thus we expect to find $2.8\times 10^{-3}\times 50=0.1$ very short microlensing events with giant sources annually (about one event during the entire OGLE-IV time span) if the probability of detection is the same for events due to free-floating planets and stars. In reality, the detection efficiency for bright events on timescales of $O(1\,\mathrm{d})$ is a factor of $2-4$ lower than for stellar events \citep{Mroz2017}. Thus, our findings support the conclusions of \citet{Mroz2017} that such Earth-mass free-floating (or wide-orbit) planets are more common than stars in the Milky Way.

\begin{acknowledgements}
P.M. acknowledges support from the Foundation for Polish Science (Program START) and the National Science Center, Poland (grant ETIUDA 2018/28/T/ST9/00096). The OGLE project has received funding from the National Science Center, Poland, grant MAESTRO 2014/14/A/ST9/00121 to A.U. 
Work by A.G. was supported by AST-1516842 from the US NSF.
I.G.S. and A.G. were supported by JPL grant 1500811.
A.G. received support from the European Research Council under the
European Union's Seventh Framework Programme (FP 7) ERC Grant Agreement n. [321035].
This research has made use of the KMTNet system operated by the Korea
Astronomy and Space Science Institute (KASI) and the data were
obtained at three host sites of CTIO in Chile, SAAO in South Africa,
and SSO in Australia.
Work by C.H. was supported by a grant (2017R1A4A1015178) from the National 
Research Foundation of Korea.
The MOA project is supported by JSPS KAKENHI Grant Numbers JSPS24253004, JSPS26247023, JSPS23340064, JSPS15H00781, and JP16H06287.
This research was supported by the I-CORE program
of the Planning and Budgeting Committee and the Israel
Science Foundation, Grant 1829/12. D.M. and A.G. acknowledge support from
the US-Israel Binational Science Foundation.
\end{acknowledgements}

\bibpunct{(}{)}{;}{a}{}{,}

\bibliographystyle{aa}
\bibliography{pap}

\end{document}